\documentclass[aps,prl,reprint,superscriptaddress,longbibliography,twocolumn]{revtex4-2}
\usepackage{graphicx} 
\usepackage{svg}
\usepackage{amsmath}
\usepackage{amssymb}
\usepackage{mathtools}
\usepackage{bbm}
\usepackage{xcolor}
\usepackage[colorlinks=true,linkcolor=blue, citecolor=green, urlcolor=blue]{hyperref}
\usepackage{multirow,booktabs}
\usepackage{youngtab}
\usepackage{soul}

\newcommand{\ket}[1]{| #1 \rangle}
\newcommand{\bra}[1]{\langle #1 |}

\begin{document}

\title{Permutation-symmetric quantum trajectories}
\date{May 2026}
\author{Elliot W. Lloyd}
\affiliation{SUPA, School of Physics and Astronomy, University of St Andrews, St Andrews, KY16 9SS, United Kingdom}
\author{Aleksandra A. Ziolkowska}
\affiliation{SUPA, School of Physics and Astronomy, University of St Andrews, St Andrews, KY16 9SS, United Kingdom}
\author{Jonathan Keeling}
\affiliation{SUPA, School of Physics and Astronomy, University of St Andrews, St Andrews, KY16 9SS, United Kingdom}

\begin{abstract}
    We show how one may perform a stochastic unravelling which respects weak permutation symmetry for general models of $N$  emitters coupled to a common system (e.g. a cavity mode). 
    Our work extends prior work by Zhang et al. [2018 New J. Phys. 20 112001] which provided an efficient quantum jump approach for a restricted subset of such problems.
    For problems involving $2$-level emitters, such an unravelling reduces the computational cost from $\mathcal{O}(N^5)$ to $\mathcal{O}(N^2)$, and with additional refinements, allows reduction to $\mathcal{O}(N)$.
    This significantly increases the range of system sizes for which one can model exact quantum dynamics of such systems.
    We further show how the method can also be applied to $d$-level systems, with computational effort scaling as $\mathcal{O}(N^{d(d-1)/2})$, and we show it allows large-$N$ simulations for $d=3$.
\end{abstract}

\maketitle

Many paradigmatic models of cavity QED involve $N$  emitters coupled to a common cavity mode.
Such models arise in contexts including simple models of lasing~\cite{haken70,scully1997quantum}, superradiant dynamics~\cite{Bonifacio1970Coherent}, superradiant phase transitions of the Dicke model~\cite{Hepp1973EquilibriumField,Wang1973PhaseSuperradiance}, and single-mode models of photon condensation~\cite{Abouelela2025stabilizing}. 
Recently there has been increasing interest in versions of such models where the emitters are more complex, with multiple electronic states, or dressed by vibrational modes~\cite{Cwik2014,Galego2015,Herrera2016Cavity}.
In the most general case, the cost of simulating such models scales exponentially with $N$, however, when the emitters are identical this cost reduces significantly.
If the Hamiltonian and all dissipators involve only sums of single emitter terms then one has strong permutation symmetry and may rewrite the model in terms of collective operators.
For two-level emitters, these are total spin operators, and the cost to simulate density-matrix evolution reduces to $\mathcal{O}(N^2 N_c^2)$ (where $N_c$ is the truncation of cavity occupation).
If there are also individual dissipation terms for the two-level emitters, the model still has weak~\cite{Buca2012,Albert2014} permutation symmetry.
Here one can use permutation symmetric density matrix approaches with a cost $\mathcal{O}(N^3 N_c^2)$. 
Such approaches have been applied to many of the models listed above~\cite{Chase2008collective,Xu2013Simulating,Richter2015,Kirton2017SuppressingDecay,Kirton2018SuperradiantModels,Shammah2018,Freter2025}.

While simulating the density matrix evolution can provide most properties of interest---i.e.\ operator expectations and correlations including multi-time correlations---there has recently been much interest in the properties that can only be found from the dynamics of individual quantum trajectories rather than their ensemble average~\cite{cao2019entanglement,li2018quantum,skinner2019measurement,choi2020quantum,gullans2020dynamical}.
Quantum trajectories~\cite{gisin1992quantum,dalibard1992wave,dum1992montecarlo,carmichael1993quantum,Plenio1998,Daley2014quantum} can be thought of as simulating individual experimental runs, and thus the state of the system conditioned on a measurement record.
Different forms of measurement (e.g. photon counting vs homodyne measurement) lead to different forms of quantum trajectory, such as the quantum jump formalism vs the quantum state diffusion equations~\cite{Plenio1998,Daley2014quantum}.
Different measurement apparatus correspond to different forms of ``stochastic unravelling'',  and thus different  entanglement within the trajectories, even though the ensemble-averaged density matrix may be the same~\cite{Vovk2022entanglement,Brown2025Gauge}.
Quantum trajectories also provide a route to cheaper numerical simulations, as one simulates a wavefunction with a size given by the Hilbert space dimension, $d_H$, rather than the density matrix size $d_H^2$.

Recent work by \citet{Macieszczak2021Quantum} has shown how abelian weak symmetries can be exploited in quantum-jump simulations. They proved that the master equation can be written in a symmetry-adapted form in which the evolution between jumps stays within a symmetry sector, while every jump takes the state to a well-defined new sector. 
A related structure had previously appeared in specific models. For example, a weak phase symmetry allows trajectory simulations where particle number has a definite value and is conserved between jumps, while individual jumps change it to another definite value~\cite{Daley2009Atomic}. 

For permutation symmetry, as discussed below, additional steps are required to realize a computationally efficient quantum-jump approach. 
We will show how these can be achieved allowing use of weak permutation symmetry for general models describing $N$ identical emitters coupled to a common cavity model.
Considering emitters as two-level systems (2LS), this approach leads to a computational cost that grows only linearly with system size, $\mathcal{O}(N N_c)$, the same scaling as would occur in the absence of individual dissipation.  
We further show that the factor $N_c$ (which typically needs to be of order $N_c \propto N$) can be eliminated.
This means a reduction from $\mathcal{O}(N^5)$ for density matrix simulations to $\mathcal{O}(N)$ for the approaches we describe here.
This enables massively larger simulations than were previously possible with individual dissipation.
As discussed further below, our work generalizes a special case that was studied in Ref.~\cite{Zhang2018MonteCarlo}.
We further show how the method can be extended beyond 2LS, to $d$-level systems, allowing large quantum simulations of more complex models.

We consider models consisting of $i=1\ldots N$ 2LS  coupled to a common system (e.g.\ a bosonic mode).
We write the general case as follows:
\begin{align}
    \hat{H}&=\sum_{i=1}^N (\hat{h}_i + \sum_\alpha \hat{x}_{i,\alpha} \hat{X}_{c,\alpha})  + \hat{H}_c,
    \label{eq:master-eqn-H}
    \\
    \partial_t \rho
    &=
    -i[\hat{H},\rho]
    + \sum_{i,\alpha} \mathcal{D}[\hat{\ell}_{i,\alpha}]
    + \sum_\alpha \mathcal{D}[\hat{C}_{\alpha}]
    \label{eq:master-eqn}
\end{align}
where $\mathcal{D}[\hat X]=\hat X \rho \hat{X}^\dagger - (1/2)\{\hat{X}^\dagger \hat{X},\rho\}$.
Here $\hat{h}_i$ is the individual Hamiltonian for the $i$th 2LS, $\hat{x}_{i,\alpha}$ is an operator of type $\alpha$ for that 2LS which couples to an operator of type $\alpha$ of the common bosonic mode $\hat{X}_{c,\alpha}$. $\hat{H}_c$ is the Hamiltonian for the common mode.  
The master equation also involves individual dissipation terms $\hat{\ell}_{i,\alpha}$ for the emitters and common bosonic mode dissipation terms $\hat{C}_{\alpha}$. 
As a concrete example (discussed further below), the Dicke model has $\hat{h}_i=\omega_0 \sigma^z_i, \hat{H}_c=\omega_c \hat{a}^\dagger \hat{a}$ and a single type of coupling $\hat{x}_{i,1}=\sigma^x_i, \hat{X}_{c,1} = (g/\sqrt{N})(\hat{a}+\hat{a}^\dagger)$.  Adding dephasing and loss for the 2LS requires two individual dissipation terms, $\hat{\ell}_{i,1} = \sqrt{\Gamma_\phi} \sigma^z_i, \hat{\ell}_{i,2}=\sqrt{\Gamma_\downarrow}\sigma^-_i$, while cavity photon loss requires $\hat{C}_{1}=\sqrt{\kappa} \hat{a}$.

The previous work by \citet{Macieszczak2021Quantum}, mentioned above, proceeded by reducing the trajectory dynamics to one symmetry sector at a time with changes of sector occurring only when jumps occur. 
Our approach extends this idea by averaging over the permutation labels, rather than simply restricting each physical trajectory to a symmetry sector. 
The averaging itself is possible due to the Schur--Weyl duality~\cite{CeccheriniSilberstein2010}, which separates the information associated with permuting the emitter labels from the collective degrees of freedom relevant to permutation-invariant dynamics.
This procedure (described below) removes the nontrivial, potentially exponentially large,  space associated with permutation of the emitter labels. 

We start from the approach to weak permutation symmetry based on collective spin states labeled by total spin $J$ and z-magnetization $M$ discussed in Refs.~\cite{Chase2008collective,Barberena2025generalized}.
As described there, one can write a permutation symmetric density matrix for the cavity-emitter system in the form:
\begin{equation}
    \rho =
    \sum_{J,M,M^\prime}
    \sum_{n_c,n_c^\prime}
    \rho^{n_c,n_c^\prime}_{J, M, M^\prime}
    \ket{n_c}\bra{n_c^\prime}
    \otimes
    \overline{\ket{J,M}\bra{J,M^\prime}}.
\end{equation}
Here $\ket{n_c}$ denotes a Fock state for the cavity;
for clarity, we will suppress the cavity state in the following discussion and restore it when summarizing our main result.
The notation for the spin represents three important points.
(1) The labels $J,M$ do not in general uniquely specify a unique state of the $N$-spin system;  one needs an additional label, which we denote $T$, to specify how the individual spins are combined to create the total spin.
Formally $T$ is associated with permutations of labels of the individual emitters, see the discussion on the $d$-level case in~\cite{SM} for details.
(2) Permutation symmetry means that the density matrix weights $\rho_{J,M,M^\prime}$ do not depend on this label $T$.
(3) Both the label $J$ and the extra label $T$ must be the same on the left and right of the density matrix.
Together these mean that
$\overline{\ket{J,M}\bra{J,M^\prime}}=(1/d_J)\sum_T 
\ket{J,T,M}\bra{J,T,M^\prime}$, where $d_J$ counts the number of states $T$ for the given $J$.  

We next summarize how the above states transform under application of the master equation.  
All contributions from the 2LS operators are given by understanding the effect of terms $\sum_i \hat X_i \rho \hat Y_i^\dagger$ (including cases with $X_i$ or $Y_i$ equal to the identity). 
It has been shown~\cite{Chase2008collective,Barberena2025generalized} that:
\begin{multline}
    \label{eq:ActionOfJumps}
    \sum_i \hat X_i \overline{\ket{J,M}\bra{J,M^\prime}} \hat Y_i^\dagger 
    =
    \sum_{\mathclap{\tau, \sigma, \sigma^\prime \in \{0,\pm 1\}}}
    f_{\hat X}(N,J,M,\tau,\sigma)
    \\
    \times
    f^\ast_{\hat Y}(N,J,M^\prime,\tau,\sigma^\prime)
    \overline{
        \ket{J+\tau,M+\sigma}
        \bra{J+\tau,M^\prime+\sigma^\prime}}
\end{multline}
The notation used here differs from \cite{Chase2008collective,Barberena2025generalized}, but is mathematically equivalent.  
In the supplement~\cite{SM} we give expressions for the functions $f_{\sigma^{\pm}},f_{\sigma^z}$ corresponding to results in Ref.~\cite{Chase2008collective,Barberena2025generalized}.
These expressions serve as a basis for any single-site spin operators.


The key observation for permutationally-symmetric stochastic unravelling is that while Eq.~\ref{eq:ActionOfJumps} involves the operator
$\overline{\ket{J,M}\bra{J,M^\prime}}$ its structure is consistent with replacing this with effective states $\ket{J,M}\bra{J,M^\prime}$.
We thus consider the dynamics of a pseudo-state
$\tilde{\rho}=\sum_{J,M,M^\prime} \rho_{J,M,M^\prime} \ket{J,M}\bra{J,M^\prime}$.
We may note that the normalization used for 
$\overline{\ket{J,M}\bra{J,M^\prime}}$ implies that $\tilde{\rho}$ will be a normalized density matrix with trace $1$ as required.
As such, $\tilde\rho$ may be used as a density matrix to calculate permutation symmetric observables: since such observables depend only on the labels $J,M$, their evolution can be simply extracted from $\tilde\rho$.  
Formally this step corresponds to representing the $S_N$ commutant by an isomorphic reduced algebra acting only on the $U(d)$ multiplicity spaces, as allowed by Schur--Weyl duality~\cite{CeccheriniSilberstein2010}.
The resulting trajectories are pure pseudo-states on these polynomial-sized spaces rather than physical pure states in exponentially large symmetry sectors.

The effect of individual dissipation on the pseudo state can be written in terms of operators defined as:
\begin{equation}
 \label{eq:defL2LS}
 \hat{L}_{\hat X,\tau} = 
 \sum_{\sigma,J,M} 
 f_{\hat X}(N,J,M,\tau,\sigma)
 \ket{J+\tau,M+\sigma} \bra{J,M},
\end{equation}
so that one can write
$\sum_i \hat{X}_i \rho \hat{Y}^\dagger_i \to \sum_\tau \hat{L}_{\hat X,\tau} \tilde{\rho} \hat{L}_{\hat{Y},\tau}^\dagger$.  
One may note that the separation of jumps with different $\tau$ ensures the density matrix remains block diagonal in the index $J$ as required.
With this form  we can then simply perform a quantum jump unravelling.  
The evolution of a trajectory pure state $\ket{\tilde{\psi}}$ over a small time $\mathrm{d}t$ is:
\begin{equation}
    \label{eq:QuantumJumpEquation}
    \ket{\tilde{\psi}} \to
    \begin{cases}
        \hat{L}_{\hat{\ell}_\alpha,\tau} \ket{\tilde{\psi}}
        /| \hat{L}_{\hat{\ell}_\alpha,\tau} \ket{\tilde{\psi}}|,
        &
        p_{\alpha,\tau} = 
        \langle \hat{L}^\dagger_{\hat{\ell}_\alpha,\tau} \hat{L}^{}_{\hat{\ell}_\alpha,\tau} \rangle \mathrm{d}t
        \\
        \hat{C}_{\alpha} \ket{\tilde{\psi}}
        /|\hat{C}_{\alpha} \ket{\tilde{\psi}}|,
        &
        p_{\alpha} = 
        \langle \hat{C}^{\dagger}_{\alpha} \hat{C}^{}_{\alpha} \rangle \mathrm{d}t
        \\
        [1- i \mathrm{d}t \hat{H}_{\text{eff}}] \ket{\tilde{\psi}},
        & \text{otherwise}
    \end{cases}
\end{equation}
where $\langle \hat O \rangle = \langle \tilde{\psi} | \hat O | \tilde{\psi} \rangle/\langle \tilde{\psi} | \tilde{\psi} \rangle$ for this unnormalized evolution. 
The first line describes individual dissipation jumps, the second line  collective dissipation jumps, and the final line the non-jump evolution. 
The effective Hamiltonian takes the form $\hat{H}_{\text{eff}}=\hat{H}-(i/2)[ \sum_{i,\alpha} \hat{\ell}^\dagger_{\alpha,i} \hat{\ell}^{}_{\alpha,i} 
+ \sum_{\alpha} \hat{C}^\dagger_{\alpha} \hat{C}^{}_{\alpha}]$.
By construction the effective Hamiltonian is permutation symmetric, and
is always block diagonal in the index $J$, i.e. 
$\hat{H}_{\text{eff}}=\sum_{J,M,M^\prime} H_{\text{eff},J,M,M^\prime}
\ket{J,M}\bra{J,M^\prime}$ (suppressing cavity indices).
This means $J$ is constant between jumps, and only changes when the 
$\hat{L}_{\hat{\ell}_\alpha,\tau}$ jumps occur, an expression of the weak permutation symmetry.
Thus, while the wavefunction $\ket{\tilde{\psi}}=
\sum_{J,M,n_c} \tilde{\psi}_{J,M,n_c} \ket{n_c}\otimes \ket{J,M}$ would naively require $\mathcal{O}(N^2 N_c)$ memory,  the evolution between jumps only needs $\mathcal{O}(N N_c)$.  

For the largest system sizes we study, the scaling of computation \emph{time} has an extra factor of $N$.
This is because at large $N$, the timestep $dt$ must reduce to ensure $p_{\alpha,\tau}$ remains small. 
Because the individual dissipation terms act on $N$ sites, their rate grows with $N$, so the required $dt$ decreases as $1/N$.
However, this only matters at very large $N$ where dissipation (rather than the Hamiltonian) fixes the required timestep.
This effect also exists in principle for the density matrix approach, but was not seen because the $N$ studied previously did not reach the limit where this dominates. 
The simulations we present used the largest step size $dt$ compatible with sufficiently approximating the exponential decay of the norm of states. 
The effective Hamiltonian step used higher-order Runge-Kutta methods to ensure smooth evolution.

\begin{figure}
    \centering
    \includegraphics[width=\columnwidth]{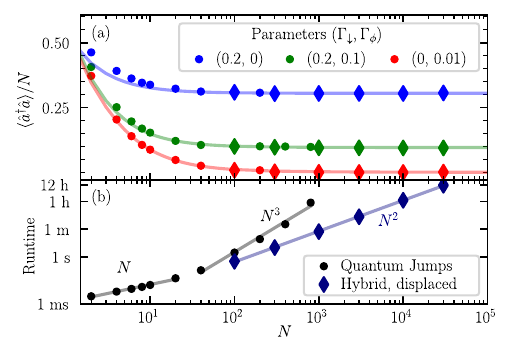}
    \caption{Dicke model with individual dephasing and dissipation.  (a) Photon number as a function of increasing system size. 
    Circles and diamonds show the results of the simple quantum jump approach vs the hybrid displaced approach described in the text.
    Solid lines compare to cumulant calculations taken from  Ref.~\cite{Kirton2017SuppressingDecay}.
    Parameters (matching Ref.~\cite{Kirton2017SuppressingDecay}) are listed in the supplement~\cite{SM}.
    Results are averaged over $n_T=2000$ to $24$ trajectories, decreasing with increasing $N$.
    (b) Run time of each trajectory simulation. 
    Circles and diamonds as described above.  The solid lines show different power law scalings.
    }
    \label{fig:dicke}
\end{figure}

To illustrate the method, Figure~\ref{fig:dicke} shows  simulations for the Dicke model with individual dephasing and loss on the atoms, i.e. the example model discussed directly after Eq.~\eqref{eq:master-eqn}. Panel (a) presents a photon number vs $N$  for system sizes up to two orders of magnitude larger than found in~\cite{Kirton2017SuppressingDecay}.
Panel (b) shows run time vs $N$ for both quantum jumps and a refined method described below.
For quantum jumps, the cavity truncation was set to $N_c=\text{max}(20, N/2)$, so at small $N$ the run time is linear in system size, while at larger $N$ one has cubic
dependence from both $N_c \propto N$ and $dt \propto 1/N$.

As noted earlier, \citet{Zhang2018MonteCarlo} discuss a special case of Eq.~\ref{eq:QuantumJumpEquation}, where the effective Hamiltonian is diagonal in the $\ket{J,M}$ basis.
In such a case the dynamics takes a particularly simple form: the trajectory is entirely determined by jumps between labels $J,M$ with no relevant dynamics between jumps.
The memory required is then independent of $N$, although the computation cost does grow with $N$ due to the increasing jump rate mentioned above.
In the supplement~\cite{SM} we show this allows direct simulation of quantum dynamics for $N\sim 10^9$.

Having established the basic unravelling approach, we next discuss two direct extensions that allow an even greater improvement in computational cost, by removing the factor of $N_c$ in the scaling noted above.

\begin{figure}
    \centering
    \includegraphics[width=\columnwidth]{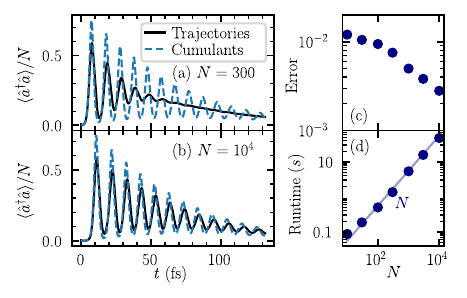}
    \caption{Superradiant emission from the Tavis-Cummings model, starting with fully excited 2LS, following~\cite{Freter2025}. 
    (a,b) Dynamics of photon number for two different values of $N$.
    Solid lines show the exact results, while dashed lines show the cumulant prediction from Ref.~\cite{Freter2025}.
    (c) Mean square error of the cumulant vs exact results over the time range shown in (a,b), showing a slow decrease with $N$.
    (d) Run time for a single trajectory vs system size. See~\cite{SM} for simulation parameters.}
    \label{fig:superradiance}
\end{figure}

One simple way to achieve this arises in models which have an additional weak $U(1)$ phase symmetry.  
In such cases, as discussed by~\citet{Daley2009Atomic,Macieszczak2021Quantum}, one will have a number operator that is conserved between the quantum jumps, and varies only when jumps occur.  
One model that has this property is the Tavis-Cummings model with photon loss and 2LS dissipation. 
This differs from the Dicke model in that it has two terms~\footnote{In the notation of Eq.~\eqref{eq:master-eqn-H} these are
$\hat{x}_{i,1}=\sigma^x_i/2, \hat{X}_{c,1} = (g/\sqrt{N})(\hat{a}+\hat{a}^\dagger)$
and
$\hat{x}_{i,2}=\sigma^y_i/2, \hat{X}_{c,2} = i (g/\sqrt{N})(\hat{a}-\hat{a}^\dagger)$.
} describing coupling between cavity and emitters: $(g/\sqrt{N})(\hat{a} \sigma^+ + \hat{a}^\dagger \sigma^-)$, but with all other terms exactly as for the Dicke model.  
Such a model has the property that between quantum jumps the total excitation number (and thus $n_c+M$) is conserved.  
As such, if one starts in a state with definite excitation number, then one can effectively eliminate the cavity state between jumps, reducing the problem to one that scales linearly with $N$. 
This model has recently been studied in Ref.~\cite{Freter2025}, to explore the effect of individual dissipation on superradiant dynamics starting from an empty cavity and all 2LS in their excited state.
In Ref.~\cite{Freter2025} permutation symmetric density matrices were combined with the block structure due to the weak $U(1)$ symmetry allowing simulations up to $N=140$.  
In Figure~\ref{fig:superradiance} we show results up to $N=10^4$. 
These results for large $N$ confirm the behavior suggested in Ref.~\cite{Freter2025}, that exact numerics approach the results of cumulants for large $N$, but approach this limit very slowly.

While the above method allows linear scaling for problems with weak $U(1)$ symmetry, this does not apply to the Dicke model.
However, an alternate approach, building on an idea discussed in Refs.~\cite{Schack1995,scully1997quantum,Muller2025nuHOPS}, can be applied. 
The idea is to consider stochastic unravellings where the phase symmetry of the photon field is broken, such that
$\langle \hat a^\dagger \hat a \rangle \simeq |\langle \hat a \rangle|^2$.
In such cases, one can significantly reduce the state space needed for the photon by making a unitary transform to displace the photon field $\hat a \to \hat a + \alpha$ such that the displaced field is centered around the vacuum state, allowing a smaller truncation $N_c$.

The quantum jump unravelling described so far does not allow for the improvement described above.
Quantum jump unravellings correspond to experiments which count photons, rather than measuring the phase of the cavity field, and thus $\langle \hat a \rangle$ remains zero throughout the quantum jump evolution.
To have a non-zero expectation of $\hat a$ one should instead consider the quantum state diffusion unravelling~\cite{gisin1992quantum}, which corresponds to a measurement record of the phase of the cavity field.
However, the simplification due to $J$ being conserved between jumps noted above relies on the use of quantum jumps.
This can be resolved, gaining both advantages simultaneously, by simulating a mixed stochastic unravelling using quantum jumps for the individual dissipation and quantum state diffusion for the cavity loss; see the supplement~\cite{SM} for details.
Figure~\ref{fig:dicke}(b) includes results showing how the run time for simulating the Dicke model scales when making use of this additional improvement.  
For the range of $N$ shown, it is possible to keep a fixed cavity truncation $N_c=6$ after this transform, so the resulting scaling is quadratic (once allowing for $dt \propto 1/N$).

\begin{figure}
    \centering
    \includegraphics[width=\columnwidth]{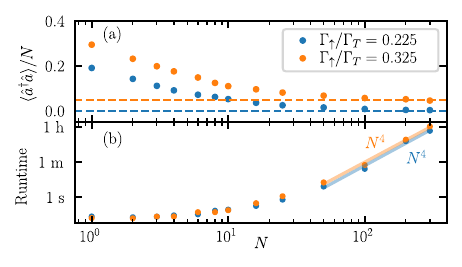}
    \caption{Inversionless lasing of a three-level system~\cite{Werren2022}
    (a) Photon number vs system size, showing convergence toward the mean-field results.  
    (b) Run time per trajectory as a function of system size.  The lines show a fit to the expected asymptotic $N^4$ scaling accounting for $dt \propto 1/N$. See~\cite{SM} for details of model and simulation parameters.}
    \label{fig:three-level}
\end{figure}

Our discussion so far has been restricted to 2LS.
However, inspired by the recent work \citet{Bastin2025Permutationally},  our results can be directly extended to $d$-level systems, as discussed in the supplement~\cite{SM}. 
Such an extension comes at the cost of more involved notation in terms of Young diagrams (labeled $\nu$) and standard Weyl tableaux (labeled $W_\nu$)~\cite{CeccheriniSilberstein2010}.  
The structure of equations for the $d$-level systems is analogous to that for the 2LS, with $\nu$ playing the role of $J$ and $W_\nu$ playing the role of $M$.
The matrix elements can be found through explicit formulae~\cite{Bastin2025Permutationally} in terms of Gelfand-Tsetlin patterns~\cite{Gelfand1950},  see~\cite{SM} for details.
As such, the computational cost of modeling the $d$-level system is given by the number of distinct Weyl tableau of a fixed shape $\nu$, and at large $N$ this asymptotically approaches $N^{d(d-1)/2}$.
The same simplifications around cavity truncation can still be applied, eliminating $N_c$ from this scaling.
This compares to $N^{d^2-1} N_c^2$ for the permutation symmetric density matrix approach.  
For a three-level system (with $N_c \propto N$) this means reduction from $N^{10}$ to $N^3$ scaling, allowing significantly larger system sizes to be reached.  
To illustrate this, Fig.~\ref{fig:three-level} shows results for a model of inversionless lasing with three-level systems introduced in Ref.~\cite{Werren2022}; the model is summarized in~\cite{SM}.
The increased system sizes accessible via trajectories confirm the behavior anticipated in Ref.~\cite{Werren2022} of mean-field results being recovered at large $N$.

The scaling of run-time with $N$ still remains prohibitive for large $d$, but simulations with $d=4,5$ and mesoscopic $N\sim30,10$ respectively should be possible, in contrast to density matrix methods where $d\leq 3$.

In summary, we have shown that one can perform stochastic unravelling in a way that respects weak permutation symmetry.
The fact that these speedups require a specific prescription for unravelling---treating the individual dissipation through mapping to effective operators $\hat{L}_{\hat X,\tau}$ and then using a quantum jump unravelling---can be seen as an example of the observation that different forms of unravelling can lead to different computational complexity~\cite{Vovk2022entanglement,Brown2025Gauge}.
The methods described here can be applied across the range of cavity-QED models describing $N$ emitters coupled to a cavity mode in contexts including lasing, photon condensation, and superradiance.
By allowing simulations to large system sizes, this allows better comparisons to approximate numerical methods such as mean-field theory, or cumulant and BBGKY expansions~\cite{Kirton2017SuppressingDecay,Kirton2018SuperradiantModels,Freter2025,muller2025hierarchical}.

One possible direction for future work is to explore how the ideas here could be extended to non-Markovian stochastic unravellings.
The Hierarchy Of Pure States (HOPS) method~\cite{Suess2014HierarchyDynamics,Muller2025nuHOPS} can be considered as a non-Markovian extension of the quantum state diffusion equation.
The key role of quantum jumps in the results above, however, suggests there could be substantial computational benefits in developing non-Markovian quantum jump formalisms, which might allow for efficient exact simulations of non-Markovian individual dissipation processes in systems of $N$ emitters.
Such exact methods could be complementary to recent approaches using 
$1/N$ expansions via the BBGKY hierarchy~\cite{muller2025hierarchical}.

A numerical library implementing the methods described in this manuscript is available at~\cite{PIMCS}.

\begin{acknowledgments}
    We acknowledge helpful discussions with Peter Kirton and Andrew Daley, and helpful comments from J.~P.~Garrahan and K.~Macieszczak on an earlier version of the manuscript.
    JMJK and AAZ acknowledge funding from EPSRC (Grant No. EP/Z533713/1).
\end{acknowledgments}

%

\onecolumngrid
\clearpage

\renewcommand{\theequation}{S\arabic{equation}}
\renewcommand{\thefigure}{S\arabic{figure}}
\setcounter{equation}{0}
\setcounter{figure}{0}
\setcounter{page}{1}

\section{Supplemental Material for: ``Permutation-symmetric quantum trajectories''}
\twocolumngrid

\section{Expressions for coefficients \texorpdfstring{$f_{\hat{X}}$}{fX}}
In the main text, we noted that the results of Refs.~\cite{Chase2008collective,Barberena2025generalized} can be written in terms of functions $f_{\hat{X}}$.  
Here we give explicit expressions for these functions for the three individual operators $\sigma^\pm,\sigma^z$.  
Any single-site operator can be written as a linear combination of these operators, so the results below form a basis for any single-site dissipation channel $\hat \ell_\alpha$.

The expressions for $f_{\hat{X}}$ were written in terms of indices $\tau,\sigma$ that defined the change of the state labels $J,M$.  
In table~\ref{tab:matrix-elements} we show the non-zero values; other combinations of $\tau,\sigma,\hat X$ give zero.
To write these formulas, we use the expressions
\begin{align}
    E_{J,N}&=\frac{1+N/2}{2J(J+1)},
    \\
    F_{J,N}&=\frac{N/2+J+2}{2(J+1)(2J+3)},
    \\
    G_{J,N}&=\frac{N/2-J+1}{2J(2J-1)}.
\end{align}

\begin{table*}[ht]
    \centering
    \begin{tabular}{l|ccc}
    $\tau$ 
    & $f_{\sigma^+}(N,J,M,\tau,+1)$
    & $f_{\sigma^-}(N,J,M,\tau,-1)$
    & $f_{\sigma^z}(N,J,M,\tau,0)$ 
    \\\hline
    $0$
    & $\sqrt{E_{J,N} (J+M+1)(J-M)}$
    & $\sqrt{E_{J,N} (J+M) (J-M+1)}$
    & $2\sqrt{E_{J,N} } M$
    \\
    $-1$
    & $-\sqrt{F_{J,N}(J-M)(J-M-1)}$
    & $\sqrt{F_{J,N}(J+M)(J+M-1)}$
    & $-2 \sqrt{F_{J,N}(J-M)(J+M)}$
    \\
    $+1$
    & $\sqrt{G_{J,N}(J+M+1)(J+M+2)}$
    & $-\sqrt{G_{J,N}(J-M+1)(J-M+2)}$
    & $-2\sqrt{G_{J,N}(J-M+1)(J+M+1)}$
    \end{tabular}
    \caption{Non-zero matrix elements for the operators $\sigma^{\pm,z}$.}
    \label{tab:matrix-elements}
\end{table*}

\section{Quantum state diffusion}

In this section we provide further details on the quantum state diffusion approach (for cavity loss) discussed in the main text. 
For reference, we summarize here the standard form of the quantum state diffusion equation~\cite{gisin1992quantum} for a Lindblad master equation with jump operators $\hat K_\alpha$.
\begin{multline}
    \ket{\psi}
    \to
    \bigg[1-i \mathrm{d}t \hat H
    -    
    \frac{dt}{2}
    \sum_\alpha 
    \left(
    \hat K^\dagger_\alpha K^{}_\alpha
    - 
    2 k_\alpha^\ast \hat K^{}_\alpha
    +
    |k_\alpha|^2 
    \right)
    \\
    +
    \sum_\alpha
    \mathrm{d} \xi_\alpha
    \left( \hat K^{}_\alpha - k_\alpha \right)
    \bigg] \ket{\psi},
\end{multline}
where $\mathrm{d}\xi_\alpha$ is a Wiener increment, satisfying
$\mathbbm{E}[\mathrm{d}\xi_\alpha]=0,\mathbbm{E}[\mathrm{d}\xi_\alpha \mathrm{d}\xi_\beta]=0,\mathbbm{E}[\mathrm{d}\xi_\alpha^\ast \mathrm{d}\xi_\beta^{}]=\delta_{\alpha\beta} \mathrm{d}t$, and $k_\alpha$ is an expectation  $k_\alpha = \langle \psi | \hat K^{}_\alpha | \psi \rangle$.

\begin{widetext}
\subsection{Hybrid quantum jump and quantum state diffusion}

The hybrid method therefore divides the jump operators into two classes: $\hat C_\alpha$ for cavity processes, and $\hat l_{i,\alpha}$ for individual processes, with $\hat L_{\hat l_\alpha}$ as the effective global operator. The individual processes will continue to be modeled by jumps, while the cavity processes will be modeled through diffusion of the state.  We may thus write:
\begin{equation}
    \ket{\psi}
    \to
    \begin{cases}
    {\hat{L}_{\hat{\ell}_\alpha,\tau} \ket{\psi}}/
    {|\hat{L}_{\hat{\ell}_\alpha,\tau} \ket{\psi}|}
    &
    p_{\alpha,\tau}=|\hat{L}_{\hat{\ell}_\alpha,\tau} \ket{\psi}|^2 \mathrm{d}t 
    \\
    \left[1-i \mathrm{d}t \hat H
    -    
    \frac{\mathrm{d}t}{2}
    \sum_{i,\alpha} \hat l^\dagger_{i,\alpha} \hat l^{}_{i,\alpha}
    -
    \frac{\mathrm{d}t}{2}
    \sum_\alpha 
    \left(
    \hat C^\dagger_\alpha C^{}_\alpha
    - 
    2 c_\alpha^\ast \hat C^{}_\alpha
    +
    |c_\alpha|^2 
    \right)
    +
    \sum_\alpha
    \mathrm{d} \xi_\alpha
    \left( \hat C^{}_\alpha - c_\alpha \right)
    \right] \ket{\psi}
    & \text{otherwise}
    \end{cases}
\end{equation}
where $c_\alpha = \langle \psi | \hat C^{}_\alpha | \psi \rangle$
Note that this expression involves an effective Hamiltonian that is modified by the effect of the individual loss processes, just as in Eq.~\ref{eq:QuantumJumpEquation}.
\end{widetext}

\subsection{Displaced operators}
In this section, we assume there is only a single cavity dissipation operator, $\hat C_1 = \sqrt{\kappa} \hat a$, as applies for all the models discussed in the main text.
When using quantum state diffusion as  above, there will in general be a non-zero expectation $c_\alpha$ for that operator.
Following Refs.~\cite{scully1997quantum,
Schack1995,Muller2025nuHOPS}, we apply a transform that removes this expectation.  
Specifically we transform $\rho \to \hat D \rho \hat D^\dagger$
with $\hat D = \exp(\alpha \hat a^\dagger - \alpha^\ast \hat a)$, which has the effect of replacing
$\hat D^\dagger \hat a \hat D = \hat a + \alpha$.
We take $\alpha$ to be time dependent, and the time derivative of $\hat D$ is then:
\begin{equation}
    \frac{\mathrm{d} \hat D}{\mathrm{d}t} 
    =
    \hat D \left( \dot{\alpha} \hat a^\dagger - \dot{\alpha}^\ast \hat a
    + \frac{1}{2}\left( \dot \alpha \alpha^\ast - \dot{\alpha}^\ast \alpha\right) \right).
\end{equation}
Substituting this into the equation of motion then allows one to find the equation of motion for $\alpha$ required in order that $\langle \hat a \rangle$ should vanish after the transformation. 
For the specific case of the Dicke model discussed in the main text this equation takes the form:
\begin{equation}
    \dot{\alpha} = ( -i \omega_c - \kappa ) \alpha
    - i \frac{g}{\sqrt{N}} 
    \langle \psi | 2 \hat{J}_x | \psi \rangle,
\end{equation}
where we have written $2\hat{J}_x=\sum_i \sigma^x_i$.
In general the equation for $\alpha$ corresponds to taking the expectation of the Heisenberg equation for $\hat a$. 
After such a transformation, the quantum state diffusion equation takes a modified form with $c_\alpha = 0$.  
Subsequently, the Hilbert space required for the cavity state can be modeled with a small truncation $N_c$, so that the computational cost of simulating the Dicke model with individual dephasing terms becomes linear in $N$.

\section{Generalization to \texorpdfstring{$d$}{d}-level systems}

In this section we summarize how the results in the main paper can be generalized to $d$-level systems.  
We make use of the approach and notation of \citet{Bastin2025Permutationally}.
As discussed there (as well as in textbooks e.g.\ Ref.~\cite{CeccheriniSilberstein2010}), for the general $d$-level case it is convenient to replace the labels $J,M$ by Young diagrams and standard Weyl tableaux.   To make our description self-contained, we briefly describe these standard results in the following sections before showing how this leads to the effective jump operators. 

\subsection{Young diagram and Weyl tableau notation}

We will use the Greek letters $\nu,\mu,\lambda$ to indicate a Young diagram representing $N$ systems with $d$ levels. 
Such a diagram consists of $N$ boxes arranged in at most $d$ rows such that the length of each successive row is not increasing.  
The diagram represents the pattern of symmetrization and antisymmetrization that construct the overall state: boxes in the same row are to be mutually symmetrized, while boxes in the same column are to be mutually antisymmetrized.  

Note that the object $\nu$ contains information equivalent to the combination of integers $J$ and $N$.
For 2LS, the index $J$ corresponds to half of the difference in length between the first and second row, 
so that maximum $J$ means all boxes are in the first row, and $J=0$ (possible only for even $N$) means equal numbers of boxes in each row.

The analog of $M$ for the $d$-level system is the standard Weyl tableau.  This object, denoted $W_\nu$, consists of placing species labels $1 \ldots d$ in the Young diagram $\nu$ such that (a) along each row the labels are not decreasing, and (b) down each column the labels are strictly increasing.    
For 2LS, the two labels $1,2$ are normally replaced by $\downarrow, \uparrow$.  The index $M$ counts half the difference between the numbers of $\uparrow$ and $\downarrow$ boxes.  The fact that the second row can only contain $\uparrow$, and that the sites above it must be $\downarrow$ encodes the standard constraint $|M| \leq J$. 

As with the two-level case, a further index, in addition to $W_\nu$ is needed to uniquely specify a quantum state.
This index (which we denoted as $T$ for the two-level case discussed in the main text)
now becomes $T_\nu$, which is a standard Young tableau.  
This consists of a way of putting each of the $N$ labels $1\ldots N$ into the boxes of the Young diagram $\nu$, such that labels in each row and each column are always increasing and  each label appears only once. 
The tableau $T_\nu$ specifies a particular way to associate the $N$ different $d$-level systems to the patterns of symmetrization and antisymmetrization.

\subsection{Density matrix equation}

As discussed in Ref.~\cite{Bastin2025Permutationally}, the $d$-level generalization of the permutation symmetry approach consists of representing the density matrix by:
\begin{align}
    \rho
    &=
    \sum_{\nu,W_\nu,W^\prime_\nu}
    \rho_{\nu,W_\nu,W^\prime_\nu}
    \overline{\ket{W_\nu}\bra{W^\prime_\nu}},
    \\
    \overline{\ket{W_\nu}\bra{W^\prime_\nu}}
    &=
    \frac{1}{d_\nu}
    \sum_{T_\nu}
    \ket{W_\nu,T_\nu}\bra{W^\prime_\nu,T_\nu}.
\end{align}
Here we have again suppressed indices describing the state of the cavity.
We note our symbol $d_\nu$ (which counts the number of Young tableaux $T_\nu$ consistent with the Young diagram $\nu$) was written as $f_\nu$ in Ref.~\cite{Bastin2025Permutationally}; we use $d_\nu$ for consistency with the two-level case in Ref.~\cite{Barberena2025generalized}.
We note also that we have chosen a normalization of the above expressions such the condition $\text{Tr}[\rho]=1$ means $\sum_{\nu,W_\nu} \rho_{\nu,W_\nu,W_\nu}=1$.
This normalization (which follows that of~\cite{Barberena2025generalized} but differs from \cite{Bastin2025Permutationally}) allows the direct interpretation the pseudo-state  $\tilde{\rho}=\sum_{\nu,W_\nu,W^\prime_\nu}
    \rho_{\nu,W_\nu,W^\prime_\nu}
    {\ket{W_\nu}\bra{W^\prime_\nu}}$ as an effective density matrix.

The action of the individual dissipation operators on such a representation can be expressed in a similar way to Eq.~\eqref{eq:ActionOfJumps} in the main text, namely:    
\begin{multline}
    \label{eq:ActionOfJumpsW}
    \sum_i \hat X_i \overline{\ket{W_\nu}\bra{W_\nu^\prime}} \hat Y_i^\dagger =
    \\
    \sum_{\mu,\lambda,W_\lambda,W^\prime_\lambda}
    f^{}_{\hat X,\mu}(W_\nu,W_\lambda)
    f^\ast_{\hat Y,\mu}(W^\prime_\nu,W^\prime_\lambda)
    \overline{\ket{W_\lambda}\bra{W_\lambda^\prime}}
\end{multline}
The change of notation simplifies the structure of this equation, but requires some explanation.
Rather than summing over changes of index, $\tau$ and $\sigma$, this expression is now written in terms of indices before and after the jump, $W_\nu \to W_\lambda$,
but the sum is restricted to those terms for which $f_{\hat X}, f_{\hat Y}$ are non-zero.  
Specifically, the diagrams $\nu,\lambda$ must be related by removing and then adding a single box at the right-hand end of a row.
There are also further constraints on $W_\nu, W_\lambda$ as discussed below.

In addition to the above points, which are just changes of notation, there is one additional feature present in this expression that was absent for the two-level case:  the sum over the index $\mu$. This sum is over the intermediate shapes of the Young diagram needed in going from diagram $\nu$ to diagram $\lambda$. 
As noted above, the allowed processes involve removing a box from one row, then adding a box to a row.  
The diagram $\mu$ is the intermediate diagram after the box has been removed.
One may note that when $\nu\neq\lambda$ there is a unique choice of intermediate diagram $\mu$.  
However when initial and final diagrams match, i.e.\  $\nu=\lambda$, there are multiple intermediate diagrams as one could remove and then replace a box in any of the occupied rows.

One can understand the structure of Eq.~\eqref{eq:ActionOfJumpsW}
through a simple physical picture.
The sum over individual dissipation operators has the effect of taking a permutation symmetric state, singling out one site, changing its label and symmetries, and then finding the overlap with final permutation symmetric states.
As discussed in detail in Ref.~\cite{Bastin2025Permutationally}, the matrix elements for that process consist of taking matrix elements (Clebsch-Gordan coefficients) between the state defined by $\ket{W_\nu}$ and the state composed of the intermediate pattern $\ket{W_\mu}\otimes\ket{s}$, where $\ket{s}$ denotes a single site in state $s$.

In the 2LS case, while a sum over intermediate states $\mu$ still formally exists, it has only two terms, and one finds that the coefficients for these terms have the same dependence on $W_\nu,W_\lambda$.
This linear dependence allows one to use a single index for the 2LS case, instead of two.

\subsection{Gelfand–Tsetlin patterns and matrix elements}

There is a simple way to enumerate the allowed changes $W_\nu \to W_\lambda$, and to caclulate their matrix elements $f^{}_{\hat X,\mu}(W_\nu,W_\lambda)$ using Gelfand–Tsetlin patterns (GT patterns)~\cite{Gelfand1950}. This is discussed in full in Ref.~\cite{Bastin2025Permutationally}; we summarize the key points here, with some minor changes in notation and normalization consistent with the normalization used above.

The GT pattern of a Weyl tableau, $G(W_\nu)$ is a triangle of numbers, growing from the bottom to the top, where each row, $k$, describes the shape (row lengths) of the filled elements of the tableau containing labels up label $k$.  
That is, the bottom row contains the shape of only the boxes with element $1$. This is a single number as this element can only exist in the top row of the tableau.  The second row up of the GT pattern  contains two numbers, describing the lengths of the two rows of the tableau that contain elements $1$ or $2$.
The top row of the GT pattern defines the shape of the overall Young diagram, with $d$ boxes stating the length of the $d$ possible rows of the diagram.
For example, one would have:
\begin{equation}
    \vcenter{\hbox{\young(1123,234,4)}}
    \to
    \left(
    \begin{array}{ccccccc}
        4&&3&&1&&0
        \\
        &4&&2&&0&
        \\
        &&3&&1&&
        \\
        &&&2&&&
    \end{array}
    \right)
\end{equation}

\subsubsection{Allowed changes}

Using these patterns, one can then write the difference between tableau via the difference of the corresponding patterns, which define which matrix elements are allowed (can be non-zero).  
Considering the sequence $W_\nu \to W_\mu \to W_\lambda$, let us consider one half of this sequence at a time.

We focus here on the first step $W_\nu \to W_\mu$ which removes a box; the same arguments apply to the reverse of the second step $W_\mu\to W_\lambda$.

The step $W_\nu \to W_\mu$ removes one box from the diagram, and has the overall effect of decreasing the population of one species, which we denote $s$, by one.
However, because of the rules for the Weyl tableau, this change may alter the (anti-)symmetries the box had with other boxes.

One may write the allowed difference of the GT patterns as 
$G(W_\nu)-G(W_\mu)=\Delta_s(\tau)$, where in addition to
$s$ labeling the species added, the pattern of changes is defined by a vector $\tau_{k=s\ldots d}$.
The pattern $\Delta_s(\tau)$ has zeros in the bottom $s-1$ rows,
and the $k$th row ($k \geq s$) contains a single 1 at position $\tau_k$ and zero elsewhere.
The set of $\Delta_s(\tau)$ defined this way define the set of allowed changes.

\begin{figure}[htpb]
    \includegraphics{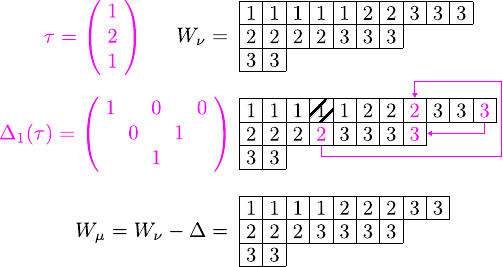}
    \caption{
    Illustration of removing a box.  
    Starting from the pattern $W_\nu$,
    The box shown in gray is removed, and other boxes moved as illustrated by arrows.
    The corresponding vector $\tau$ and GT pattern change $\Delta_{1}(\tau)$ are shown to the left.    
    }
    \label{fig:BoxRemoval}
\end{figure}

Figure~\ref{fig:BoxRemoval} shows a $\Delta_1$ pattern that removes a box with label 1, breaking anti-symmetrized and symmetrized pairings with boxes with labels 2 and 3 respectively, and constructs an anti-symmetrized pair from these boxes. The overall symmetry change is determined by the intermediate row changes from the GT pattern.

The final element of the vector $\tau_d$, which determines the top row of $\Delta_s(\tau)$, plays a special role.  
This defines the change of shape of the overall Young diagram, defining which row of the diagram gains an extra box.
See Ref.~\cite{Bastin2025Permutationally} for further discussion.

The set of allowed changes $W_\nu \to W_\lambda$ are those such that there exists an intermediate $W_\mu$ such that both
$G(W_\nu)-G(W_\mu)$ and $G(W_\lambda)-G(W_\mu)$ are allowed change patterns.  That means that allowed changes require
\begin{equation}
    G(W_\lambda) - G(W_\nu) = \Delta_{s}(\tau) - \Delta_{s^\prime}(\tau^\prime)
\end{equation}
for some $s,s^\prime,\tau,\tau^\prime$.
Such a change involves a species change from $s^\prime \to s$.

\subsubsection{Matrix elements}

Following~\cite{Bastin2025Permutationally}, the matrix elements can be written in terms of GT patterns by defining the operator $\hat{g}^{(W_\nu,W_\lambda)}_\mu$, written below, which acts on the space of a single $d$-level system.
In terms of these operators, the matrix elements of allowed
transitions take the form:
\begin{equation}
    \label{eq:deffgeneral}
    f^{}_{\hat X,\mu}(W_\nu,W_\lambda)
    =
    \sqrt{N \frac{d_\mu}{d_\nu}}
    \text{Tr}\left[ \hat{g}^{\dagger(W_\nu,W_\lambda)} \hat{X} \right]
\end{equation}
Note that the prefactor here differs from that in~\cite{Bastin2025Permutationally} due to our changed normalisation.
The operator $\hat{g}^{(W_\nu,W_\lambda)}$  takes the form
\begin{equation}
    \label{eq:defgop}
    \hat{g}^{(W_\nu,W_\lambda)}_\mu
    =
    \sum_{W_\mu}
    \zeta^{(W_\lambda)}_{s,\tau}
    \zeta^{(W_\nu)}_{s^\prime,\tau^\prime}
    \ket{s}\bra{s^\prime},
\end{equation}
where $s,\tau$ and $s^\prime,\tau^\prime$ are the labels required such that $G(W_\lambda)-G(W_\mu)=\Delta_s(\tau)$
and $G(W_\nu)-G(W_\mu)=\Delta_{s^\prime}(\tau^\prime)$ respectively.  
To complete the definition the coefficients $ \zeta^{(W_\lambda)}_{s,\tau}$ are written in terms of elements of the GT pattern of the the intermediate tableau,
$W_\mu=W_\lambda-\Delta_s(\tau)$, written as 
$p_{j,k}=[G(W_\mu)]_{j,k} + k -j$.
Indexing of $[G(W_\mu)]_{j,k}$ works from left to right (index j) and bottom to top (index k).
One then has:
\begin{align}
    \zeta^{(W_\lambda)}_{s,\tau}
    &=
    \Bigg[
    \frac{\prod_{k=1}^s (p_{\tau_{s+1},s+1}-p_{k,s})}
        {\prod_{\substack{k=1\quad\;\\k\neq \tau_{s+1}}}^{s+1} 
        \!\!\!\!\!\! (p_{\tau_{s+1},s+1}-p_{k,s+1})}
    \Bigg]^{1/2} \!\!\!
    \prod_{l=s+2}^d A_l,
    \nonumber\\
    A_l
    &= \text{sgn}(\tau_{l-1}-\tau_l)
    \Bigg[
    \prod_{\substack{k=1\\k\neq \tau_{l}}}^{l}
    \frac{p_{\tau_{l-1},l-1}-p_{k,l}+1}{p_{\tau_l,l}-p_{k,l}}
    \nonumber\\
    &\qquad\times
    \prod_{\substack{k=1\\k\neq \tau_{l-1}}}^{l-1}
    \frac{p_{\tau_{l},l}-p_{k,l-1}}{p_{\tau_{l-1},l-1}-p_{k,l-1}+1}
    \Bigg]^{1/2}.
\end{align}

To provide some intuition for Eq.~\eqref{eq:deffgeneral}, one may note that in cases where $\hat X = \ket{s^\prime}\bra{s}$, one can immediately read off the species changes $s,s^\prime$ in the trace in Eq.~\eqref{eq:defgop}.
This is analogous to the statement for the 2LS in table~\ref{tab:matrix-elements} that for $\hat X = \sigma^\pm$, only a single change $M\to M\pm 1$ is allowed.  
In the case $\hat X=\mathbbm{1}$ one can also show that
$f^{}_{\mathbbm{1},\mu}(W_\nu,W_\lambda)=\delta_{W_\nu,W_\lambda}$ as expected.

\subsection{Stochastic unravelling}

As with the 2LS case, stochastic unravelling can occur by replacing the full density matrix evolution by an equivalent dynamics of a pseudo-state $\tilde{\rho}=\sum_{\nu,W_\nu,W^\prime_\nu}  \rho_{\nu,W_\nu,W^\prime_\nu} \ket{W_\nu}\bra{W^\prime_\nu}$.
As for the 2LS, one then seeks to find effective jump operators that act on this pseudo-state to allow stochastic unravelling.
Because of the additional sum over intermediate diagrams $\mu$, there is an additional complication in translating the general problem: a single dissipative process may now lead to more than one effective operator.
That means that in place of the single sum over total symmetry change $\tau$ that appeared in Eq.~\eqref{eq:defL2LS} for 2LS, we must now label operators by the two changes $\nu \to \mu$ and $\mu \to \lambda$.
To be precise, the label(s) playing the role of $\tau$ for the two-level case are in fact the final elements of the vectors $\tau$ appearing in $\Delta_s(\tau)$ for the two transitions.  
These values determine the row that loses a box in the transition $\nu\to\mu$ and the row that gains a box for $\mu\to\lambda$.
We will use $(\nu-\mu,\lambda-\mu)$ below to label this: one may consider this label as a pair of numbers labelling the rows with boxes removed and added.

Using the above, one can write
\begin{equation}
    \hat{L}_{\hat X,(\nu-\mu,\lambda-\mu)}^{}
    =
    \sum_\nu
    \sum_{W_\nu,W_\lambda}
    f^{}_{\hat X,\mu}(W_\nu,W_\lambda)
    \ket{W_\lambda}\bra{W_\nu},
\end{equation}
The sum over $\nu$ appearing here plays the same role as the sum over $J$ appearing in Eq.~\ref{eq:defL2LS}, while the sums over
$W_\nu,W_\lambda$ here are equivalent to the sums over $M,M+\sigma$ in Eq.~\ref{eq:defL2LS}.
In terms of the above operator, we then have:
\begin{equation}
    \sum_i \hat X^{}_i \rho \hat Y^\dagger_i
    \to
    \sum_{(\nu-\mu,\lambda-\mu)}
    \hat{L}_{\hat X,(\nu-\mu,\lambda-\mu)}^{}
    \tilde{\rho}
    \hat{L}_{\hat Y,(\nu-\mu,\lambda-\mu)}^{\dagger}.
\end{equation}
As noted above, when $\nu \neq \lambda$, $\mu$ is uniquely determined, so the addition of the label $\mu$ is only important when $\nu=\lambda$.
We observe that for the cases explored numerically for $d=2,3$, it appears that of the $d$ distinct terms $\mu$ possible, there seem only to be $d-1$ linearly independent operators
$\hat{L}_{\hat X,(\nu-\mu,\lambda-\mu)}^{}$, a generalization of the statement that for $d=2$ there is only one such operator.

Once these collective operators are defined, the quantum jump formalism takes a similar form to Eq.~\eqref{eq:QuantumJumpEquation}, with the only change being the additional label enumerating the jump operators.

\section{Details of models studied}

In this section we summarize the parameters and details of the models shown as illustrations in the main text.

\subsection{Dicke model}

Figure~\ref{fig:dicke} shows the long-time steady state photon number for the Dicke model with individual dephasing and loss as described in the text.  Table~\ref{tab:dicke-params} shows the parameters used for this model; these are the same as those used in Ref.~\cite{Kirton2017SuppressingDecay}.  For $\Gamma_\downarrow, \Gamma_\phi$, the values used for these parameters are either zero or non-zero as labeled in the legend of Fig.~\ref{fig:dicke}.

\begin{table}[h]
    \centering
    \begin{tabular}{c|c|c|c|c|c|c}
 Parameter & $\omega_0$ & $\omega_c$ & $g$\footnote{Ref.~\cite{Kirton2017SuppressingDecay} defines this as $g\sqrt{N}$, while we have written $g/\sqrt{N}$ in the Hamiltonian.} & $\kappa$ & $\Gamma_\downarrow$ & $\Gamma_\phi$ \\\hline
Value(s) & 0.5 & 1 & 0.9 & 1 & $\{0,0.2\}$ & $\{0,0.1\}$
\end{tabular}
    \caption{(Dimensionless) parameters used for the simulations in Fig.~\ref{fig:dicke}.}
    \label{tab:dicke-params}
\end{table}

\subsection{Ringing superradiance of the Tavis-Cummings model}

Figure~\ref{fig:superradiance} showed the time evolution for the Tavis-Cummings model with individual dephasing and loss as described in the text.  Table~\ref{tab:superradiance-params} shows the parameters used for this model; these  are the same as those used in Ref.~\cite{Freter2025}.
The initial state is prepared with no photons and all 2LS in their excited state.  

\begin{table}[h]
    \centering
    \begin{tabular}{c|c|c|c|c|c}
 Parameter & $\omega_c-2\omega_0$\footnote{Ref.~\cite{Freter2025} defines $\omega_0/2$ as the prefactor of $\sigma^z$, introducing a factor of two appearing here.}
 & $g$\footnote{Ref.~\cite{Freter2025} defines this as $g\sqrt{N}$, while we have written $g/\sqrt{N}$ in the Hamiltonian.}
 & $\kappa$
 & $\Gamma_\downarrow$
 & $\Gamma_\phi$ \\\hline
Value (eV)  & -0.35 & 0.4 & 0.01 & 0.0001 & 0.0075
\end{tabular}
    \caption{Parameters used for the simulations in Fig.~\ref{fig:superradiance}}
    \label{tab:superradiance-params}
\end{table}

\subsection{Three-level lasing model}

Figure~\ref{fig:three-level} showed results for a three-level lasing model introduced in Ref.~\cite{Werren2022}.  
We will write this in terms of operators $\sigma^{\alpha\beta}=\ket{\alpha}\bra{\beta}$ with $\alpha,\beta\in\{1,2,3\}$.  
In terms of such operators, the terms in our Eqs.~\eqref{eq:master-eqn-H},\eqref{eq:master-eqn} take the form:
\begin{align*}
    \hat{h}_i &= \omega_e \sigma^{33}_i + (\Omega \sigma^{21}_i
    + \text{H.c.}), 
    &
    \hat{H}_c&=\omega_c \hat a^\dagger \hat a, 
    \\
    \hat{x}_{i,1}&=\frac{g}{\sqrt{N}} (\sigma^{31} + \sigma^{31}),
    &
    \hat{X}_{c,1}&=\frac{1}{2}(\hat a + \hat a^\dagger),
    \\
    \hat{x}_{i,2}&=\frac{ig}{\sqrt{N}} (\sigma^{13} - \sigma^{31}),
    &
    \hat{X}_{c,2}&=\frac{i}{2}(\hat a - \hat a^\dagger),
    \\
    \hat{x}_{i,3}&=\frac{g}{\sqrt{N}} (\sigma^{32} + \sigma^{32}),
    &
    \hat{X}_{c,3}&=\frac{1}{2}(\hat a + \hat a^\dagger),
    \\
    \hat{x}_{i,4}&=\frac{ig}{\sqrt{N}} (\sigma^{23} - \sigma^{32}),
    &
    \hat{X}_{c,4}&=\frac{i}{2}(\hat a - \hat a^\dagger).
\end{align*}
This describes Jaynes-Cummings type coupling of the cavity to both the $1\leftrightarrow 3$ and $2 \leftrightarrow 3$ transitions, along with coherent pumping of the $1 \leftrightarrow 2$ transition (where $\Omega$ may be complex).
The forms of dissipation considered are cavity loss $\hat{C}=\sqrt{\kappa} \hat a$, and four individual pumping and decay terms:
\begin{align*}
    \hat{\ell}_{i,1}&=\sqrt{\Gamma_\uparrow} \sigma^{31}_i 
    &
    \hat{\ell}_{i,2}&=\sqrt{\Gamma_\uparrow} \sigma^{32}_i 
    \\
    \hat{\ell}_{i,3}&=\sqrt{\Gamma_\downarrow} \sigma^{13}_i 
    &
    \hat{\ell}_{i,4}&=\sqrt{\Gamma_\downarrow} \sigma^{23}_i 
\end{align*}
Table~\ref{tab:three-level-params} provides the parameter values used.

\begin{table}[h]
    \centering
    \begin{tabular}{c|c|c|c|c|c}
 Parameter & $\omega_e-\omega_c$  & $g$ & $\kappa$ & $\Gamma_T$ \\\hline
Value & $\Omega$ & $0.9\Omega$ & $0.8\Omega$ & $\Omega$
\end{tabular}
    \caption{Dimensionless parameters (in units of $\Omega$) used for the simulations in Fig.~\ref{fig:three-level}, where $\Gamma_T = 2(\Gamma_\uparrow + \Gamma_\downarrow)$.}
    \label{tab:three-level-params}
\end{table}

The results shown in the text show the long-time value of the photon field. 
These are compared to cumulant results taken from Ref.~\cite{Werren2022}.

\subsection{Special case of \citet{Zhang2018MonteCarlo}}

\begin{figure}
    \centering
    \includegraphics[width=\columnwidth]{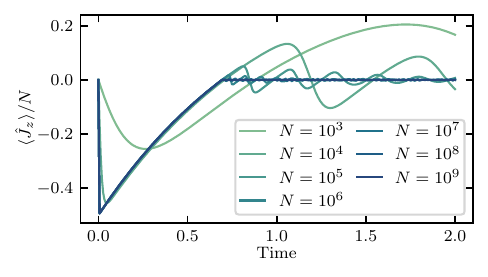}
    \caption{Spin-only model where $\hat{H} = \hat{J}_z$, with pumping $\Gamma_\uparrow = 1$ for $\Gamma_\uparrow \sum_i \mathcal{D}[\sigma_i^+]$ and collective emission $\Gamma_c = 0.01$ for $\Gamma_c \mathcal{D}[\sum_i \sigma_i^-]$. The initial state is $\ket{J=N/2, M=0}$.}
    \label{fig:diagonal-heff}
\end{figure}

As mentioned in the main text, \citet{Zhang2018MonteCarlo} discuss a special case which corresponds to the effective Hamiltonian being diagonal in the collective spin representation.
Here we briefly present some results showing what can be achieved in this special case where the state can be described entirely by the two indices $J,M$, so the required memory becomes $\mathcal{O}(1)$, independent of $N$,
as the evolution under $\hat{H}_{\textrm{eff}}$ becomes a trivial phase rotation between jumps.
The computational complexity becomes $\mathcal{O}(N)$, due to the increasing jump rate at large $N$ as noted in the main text.
This allows one to simulate even larger system sizes, $N = 10^9$, as shown in Figure~\ref{fig:diagonal-heff}.

\citet{Zhang2018MonteCarlo} discuss only the case of 2-level systems.
For $d$-level systems with diagonal $H_{\text{eff}}$ there could still be a speedup associated with replacing the evolution between jumps with a simple phase accumulation.
However, in general, collective jump operators for $d$-level systems can create superposisitions of different Weyl Tableaux, so the extension of the model discussed here to $d>2$ is not so simple.

\subsection{Hardware used for simulations}

The results presented are from trajectories using 24 cores of an AMD EPYC 9634 CPU, provided by the University of St Andrews ``Hypatia'' HPC cluster.

\end{document}